\documentclass[preprint,authoryear,12pt]{elsarticle}

\usepackage{amssymb}
\usepackage{amsthm}

\journal{Journal of Symbolic Computation}

\usepackage{xspace}
\usepackage{amsmath}
\usepackage[lined,ruled,vlined,linesnumbered]{algorithm2e}
\SetKwInput{KwEnsures}{Ensures}
\usepackage{color}
\usepackage{lscape}
\usepackage{booktabs}

\newcommand{\colechelon}{\texttt{ColEchTrans}\xspace}
\newcommand{\cupd}{\texttt{CUP}\xspace}
\newcommand{\gauss}{\texttt{Gauss}\xspace}
\newcommand{\gaussjordan}{\texttt{GaussJordan}\xspace}
\newcommand{\lqup}{\texttt{LQUP}\xspace}
\newcommand{\lsp}{\texttt{LSP}\xspace}
\newcommand{\lu}{\texttt{LU}\xspace}
\newcommand{\lup}{\texttt{LUP}\xspace}
\newcommand{\mm}{\texttt{MM}\xspace}
\newcommand{\ple}{\texttt{PLE}\xspace}
\newcommand{\plu}{\texttt{PLU}\xspace}
\newcommand{\pluq}{\texttt{PLUQ}\xspace}
\newcommand{\qlup}{\texttt{QLUP}\xspace}
\newcommand{\redcolechelon}{\texttt{RedColEchTrans}\xspace}
\newcommand{\trmm}{\texttt{TRMM}\xspace}
\newcommand{\trlum}{\texttt{TRLUM}\xspace}
\newcommand{\trsm}{\texttt{TRSM}\xspace}
\newcommand{\trtri}{\texttt{TRTRI}\xspace}
\newcommand{\trulm}{\texttt{TRULM}\xspace}

\newcommand{\Z}{\ensuremath{\mathbb{Z}}\xspace}
\newcommand{\Q}{\ensuremath{\mathbb{Q}}\xspace}
\newcommand{\GO}[1]{\ensuremath{O(#1)}\xspace}

\newcommand{\submatrix}[5]{\ensuremath{#1_{#2..#3}^{#4..#5} \xspace}}
\newcommand{\rowofmatrix}[4]{\ensuremath{#1_{#2}^{#3..#4} \xspace}}
\newcommand{\colofmatrix}[4]{\ensuremath{#1_{#2..#3}^{#4} \xspace}}
\newcommand{\interval}[2]{\ensuremath{{#1..#2} \xspace}}

\theoremstyle{definition}
\newtheorem{defn}{Definition}
\newtheorem{examp}{Example}
\newtheorem{prop}{Proposition} 
\newproof{pf}{Proof}

\renewcommand{\arraystretch}{.7}
\begin{document}

\begin{frontmatter}



\title{Rank-profile revealing Gaussian elimination\\ and the CUP matrix decomposition}

\author[lip]{Claude-Pierre Jeannerod}
\ead{claude-pierre.jeannerod@ens-lyon.fr}

\author[lig]{Cl\'ement Pernet}
\ead{clement.pernet@imag.fr}

\author[waterloo]{Arne Storjohann}
\ead{astorjoh@uwaterloo.ca}

\address[lip]{
  INRIA,
  Laboratoire LIP (CNRS, ENS de Lyon, INRIA, UCBL), Universit\'e de Lyon,
  46, all\'ee d'Italie, F-69364 Lyon Cedex 07, France}

\address[lig]{
  INRIA, Laboratoire LIG (CNRS, Grenoble INP, INRIA, UJF, UPMF)
  ENSIMAG Antenne de Montbonnot,
  51, ave Jean Kuntzmann,
  F-38330 Montbonnot Saint-Martin, France
}

\address[waterloo]{
  David R.~Cheriton School of Computer Science,
  University of Waterloo,
  Ontario, Canada, N2L 3G1
}

\begin{abstract}
Transforming a matrix over a field to echelon form, or decomposing the
matrix as a product of structured matrices that reveal the rank profile,
is a fundamental building block of computational exact linear algebra.
This paper surveys the well known variations of such decompositions and
transformations that have been proposed in the literature.  We present
an algorithm to compute the \cupd  decomposition of a matrix, adapted
from the \lsp algorithm of Ibarra, Moran and Hui (1982), and show
reductions from the other most common Gaussian elimination based
matrix transformations and decompositions to the \cupd decomposition.
We discuss the advantages of the \cupd algorithm over other existing
algorithms by studying time and space complexities: the asymptotic
time complexity is rank sensitive, and comparing the constants of the
leading terms, the algorithms for computing matrix invariants based on
the \cupd decomposition are always at least as good except in one case.
We also show that the \cupd algorithm, as well as the computation of
other invariants such as transformation to reduced column echelon form
using the \cupd algorithm, all work in place, allowing for example to
compute the inverse of a matrix on the same storage as the input matrix.
\end{abstract}


\end{frontmatter}


\section{Introduction}

Gaussian elimination and the corresponding matrix decompositions, such as
the \lu decomposition $A=LU$ as the product of lower triangular $L$ and
upper triangular $U$, are fundamental building blocks in computational
linear algebra that are used to solve problems such as linear systems,
computing the rank, the determinant and a basis of the nullspace of
a matrix.  The \lu decomposition, which is defined for matrices whose
leading principal minors are all nonsingular, can be generalized to
all nonsingular matrices by introducing pivoting on one side (rows or
columns), leading to the \lup or \plu decomposition, $P$ a permutation
matrix.  Matrices with arbitrary dimensions and rank can be handled by
introducing pivoting on both sides, leading to the \lqup decomposition of
\citet{IbMoHu82} or the \pluq decomposition \citep{GoVa96,Jeffrey:2010},
$Q$ a second permutation matrix.  We recall the precise definitions
of these decompositions in Section~\ref{sec:defi}.  For now, we note
that the decompositions are not unique.  In particular, they can differ
depending on the pivoting strategy used to form the permutations $P$
and $Q$.  Whereas in numerical linear algebra \citep{GoVa96} pivoting
is used to ensure a good numerical stability, good data locality, and
reduce the fill-in, the role of pivoting differs in the context of
exact linear algebra.  Indeed, only certain pivoting strategies for
these decompositions will reveal the rank profile of the matrix (see
Section~\ref{sec:defi}), which is crucial information in many applications
using exact Gaussian elimination, such as Gr\"obner basis computations
\citep{F99a} and computational number theory~\citep{stein2007modular}.

In this article we consider matrices over an arbitrary field and analyze
algorithms by counting the required number of arithmetic operations from
the field.  Many computations over \Z or \Q  (including polynomials with
coefficients from these rings) reduce to linear algebra over finite
fields using techniques such as linearization or homomorphic imaging
(e.g., reduction modulo a single prime, multi-modular reduction,
$p$-adic lifting).  Applications such as cryptanalysis make intensive
use of linear algebra over finite fields directly.  Unlike arithmetic
operations involving integers, the cost of an operation in a finite
field does not depend on the value of its operands, and counting the
number of field operations is indicative of the actual running time.

The asymptotic time complexity of linear algebra of matrices over a field
has been well studied.  While Gaussian elimination can be performed
on an $n\times n$ matrix using $\GO{n^3}$ arithmetic operations, a
subcubic complexity of $\GO{n^\omega}$ with $\omega<2.38$ the exponent
of matrix multiplication \citep[see][\S 12.1]{vzGG99} is obtained using
\citeauthor*{BuHo74}'s \citeyearpar{BuHo74} block recursive algorithm
for the \lup decomposition.  Algorithms for computing the \lsp and \lqup
decompositions of rank deficient matrices are given by \citet{IbMoHu82}.
As an alternative to decomposition, \citet{KelGeh85} adapts some earlier
work by \citet{Sch73} to get an algorithm that computes a nonsingular
transformation matrix $X$ such that $AX$ is in column echelon form.
For an $m\times n$ matrix such that $m \le n$, all of the Gaussian
elimination based algorithms just mentioned have an $\GO{nm^{\omega-1}}$
time complexity.

Only recently has the complexity of computing a rank profile revealing
decomposition or a transformation to echelon form been studied in
more details: algorithms with a rank-sensitive time complexity of
$\GO{mnr^{\omega-2}}$ for computing a transformation to echelon and
reduced echelon forms are given by \citet{StoMul98} and \citet{Sto00};
similar rank-sensitive time complexities for the \lsp and \lqup
decompositions of \citet{IbMoHu82} have been obtained by \citet{Jea06}.
On the one hand, while offering a rank-sensitive time complexity, the
algorithms in \citep{StoMul98,Sto00,Jea06} are not necessarily in-place.
On the other hand, \citet{DuGiPe08} describe an in-place variant of the
\lqup decomposition but, while analyzing the constant in the leading term,
does not achieve a rank-sensitive complexity.

The aim of this article is to gather, generalize, and extend these
more refined analyses to the most common Gaussian elimination based
matrix decompositions.  We propose an algorithm computing a new matrix
decomposition, the \cupd decomposition, and show reductions to it
for all common matrix decompositions.  We assess that, among all other
matrix decompositions, the algorithm for \cupd is the preferred Gaussian
elimination algorithm to be used as a building block routine for the
following reasons.

\begin{enumerate}
\item We show how all other decompositions can be recovered from
the \cupd decomposition. Furthermore, the time complexities for all
algorithms computing the alternative decompositions and transformations
(considering the constant in the leading term) are never better
than the proposed reduction to \cupd, except by a slight amount in
one case.
\item The complexity of the algorithm for \cupd is rank sensitive
(whereas no such result could be produced with some of the other
algorithms).
\item The reduction to \cupd allows us to perform these computations
in-place, that is, with essentially no more memory than what the
matrix products involved already use.
\item The \cupd decomposition offers the best modularity: combined with
other classic routines like \trsm, \trtri, \trulm and \trlum (which
shall be recalled in the course of the article), it allows to compute
solutions to other linear algebra problems such as rank, rank profile,
nullspace, determinant and inverse  with the best time complexities.
\end{enumerate}

\subsection{Notation and definitions}

Matrix entries are accessed using zero-based indexing: for example,
$a_{0,1}$ denotes the entry lying on the first row and second column
of the matrix $A=[a_{i,j}]$.  In order to make the description of our
algorithms simpler, we adopt the following convention \citep{Python}:
intervals of indices include the lower bound  but exclude the upper bound,
that is,
\[
\interval{a}{b} = \{a,a+1,\dots,b-1\}.
\]
Our algorithms make heavy use of conformal block decompositions
of matrices. In the literature, new variable names are typically
used for each block. Instead, we refer to blocks using intervals
of indices in order to emphasize the fact that no memory is being
allocated and that our algorithms actually work in-place.  The notation
$\submatrix{A}{a}{b}{c}{d}$ represents the submatrix of $A$ formed by
the intersection of rows of index from $a$ to $b-1$, and the columns of
index from $c$ to $d-1$.  For example, a $2\times 2$ block decomposition
of an $m \times n$ matrix $A$ shall be referred to as
\[
A=
\renewcommand{\arraystretch}{1.25}
\left[\begin{array}{c|c}
  \submatrix{A}{0}{k}{0}{\ell} &  \submatrix{A}{0}{k}{\ell}{n} \\
\hline
  \submatrix{A}{k}{m}{0}{\ell} &  \submatrix{A}{k}{m}{\ell}{n} \\
\end{array}\right]
\]
for suitable integers $k$ and $\ell$.  Similarly,
$\rowofmatrix{A}{a}{c}{d}$ denotes the subrow of row $a$ of $A$ whose
column indices range from $c$ to $d-1$.

Following \citet{IbMoHu82}, we say a rectangular matrix $A=[a_{i,j}]$
is {\it upper triangular} if $a_{i,j} = 0$ for $i>j$, and that it is
{\it lower triangular} if its transpose $A^T$ is upper triangular.
A matrix that is either upper or lower triangular is simply called
{\it triangular}, and if in addition $a_{i,i} = 1$ for all $i$ then
it is called {\it unit triangular}.  For two $m\times n$ matrices $L$
and $U$ such that $L=[\ell_{i,j}]$ is unit lower triangular and $U =
[u_{i,j}]$ is upper triangular, we shall use the notation $[L\backslash
U]$ to express the fact that $L$ and $U$ are stored one next to the other
within the same $m\times n$ matrix.  Thus, $A = [L\backslash U]$ is the
$m \times n$ matrix $A = [a_{i,j}]$ such that $a_{i,j} = \ell_{i,j}$
if $i>j$, and $a_{i,j} = u_{i,j}$ otherwise.

For a permutation $\sigma:\{0,\dots ,n-1\}\rightarrow\{0,\dots, n-1\}$,
the associated permutation matrix $P=[P_{i,j}]_{0 \le i,j <n}$ is
defined by $P_{i,\sigma(i)}=1$ and $P_{i,j}=0$ for $j\neq \sigma(i)$.
Multiplying a matrix on the left by $P$ applies the permutation $\sigma$
to the rows of that matrix, while multiplying on the right by the
transpose $P^T = P^{-1}$ applies $\sigma$ to its columns. We denote
by $T_{i,j}$ the permutation matrix that swaps indices $i$ and $j$
and leaves the other elements unchanged.

\subsection{Organization of the article}
Section~\ref{sec:defi} reviews and exposes the links between the most
common matrix decompositions originating from Gaussian elimination: \lu,
\lup, \texttt{Turing\-Decom\-po\-sition}, \lsp, \lqup, and \qlup, together
with our variant of the \lsp decomposition, the \cupd decomposition.
Section~\ref{sec:algo} gives algorithms to compute all these matrix
decompositions with no extra memory allocation.  More precisely, we
give an algorithm for computing a \cupd decomposition and show how all
of the other decompositions can be derived from it.  The advantages of
the \cupd algorithm compared to other existing algorithms for Gaussian
elimination are discussed in Section~\ref{sec:discussion}, based on an
analysis of time and space complexities.  In Section~\ref{sec:rowech}
we comment on the \ple decomposition, which is the row-counterpart of
the \cupd decomposition, and we conclude in Section~\ref{sec:conclusion}.

\section{Review of Gaussian elimination based matrix decompositions}
\label{sec:defi}
Throughout this section, let $A$ be an $m \times n$ matrix with rank $r$.
The {\it row rank profile} of $A$ is the lexicographically smallest
sequence of $r$ row indices $i_0<i_1 <\dots<i_{r-1}$ such that the
corresponding rows of $A$ are linearly independent.  The matrix $A$
is said to have {\it generic rank profile} if its first $r$ leading
principal minors are nonzero.

\subsection{\lu based decompositions}

If $A$ has generic rank profile it has a unique \lu decomposition: $A =
LU$ for $L$ an $m \times m$ unit lower triangular matrix with last $m-r$
columns those of the identity, and $U$ an $m \times n$ upper triangular
matrix with last $m-r$ rows equal to zero.  In our examples, zero entries
of a matrix are simply left blank, possibly nonzero entries are indicated
with $*$, and necessarily nonzero entries with $\overline{*}$.

\begin{examp}
The \lu decomposition of a $6\times 4$ matrix $A$ with rank 3.
$$
 \overset{A}{\begin{bmatrix}
     *&*&*&*\\
     *&*&*&*\\
     *&*&*&*\\
     *&*&*&*\\
     *&*&*&*\\
     *&*&*&*\\
   \end{bmatrix}}
=
 \overset{L}{\begin{bmatrix}
     1&&&\\
     *&1&&\\
     *&*&1&\\
     *&*&*&1\\
     *&*&*& &1\\
     *&*&*& & &1\\
   \end{bmatrix}}
 \overset{U}{\begin{bmatrix}
    \overline{*}&*&*&*\\
    &\overline{*}&*&*\\
    &&\overline{*}&*\\
    \\ \\ \\
   \end{bmatrix}}
$$
\end{examp}
Uniqueness of $U$ and the first $r$ rows of $L$ follows from the generic
rank profile condition on $A$, while uniqueness of the last $m-r$ rows
of $L$ follows from the condition that $L$ has last $m-r$ columns those
of the identity matrix.  If $A$ has row rank profile $[0,1,\ldots,r-1]$,
then it only takes a column permutation to achieve generic rank profile;
for such input matrices, \citeauthor{AhHoUl74} (\citeyear{AhHoUl74},
\S 6.4) and \citet{BuHo74} give a reduction to matrix multiplication
for computing an \lup decomposition: $A = LU P$, with $P$ a permutation
matrix and $AP^T = LU$ the \lu decomposition of $AP^T$.  Allowing row
and column permutations extends the \lu decomposition to any matrix,
without restriction on the rank profile.  In the context of numerical
computation, the row and column permutations leading to the best
numerical stability are chosen \citep[see for example][\S 3.4]{GoVa96},
and are referred to as partial pivoting (e.g., column permutations only)
and complete pivoting (column and row permutations). In the context of
symbolic computation, a key invariant of the input matrix that should be
revealed is the rank profile.  In the next subsection we recall the most
common matrix decompositions based on the column echelon form of $A$,
thus all revealing the row rank profile of $A$.

\subsection{Rank profile revealing decompositions}
\label{ssec:rprevealdecomp}
Let $A$ be an $m \times n$ input matrix with row rank profile
$[i_0,\dots,i_{r-1}]$. Recall how the iterative version of the Gaussian
elimination algorithm transforms $A$ to column echelon form. The algorithm
detects the row rank profile of $A$ during the elimination. For $i =
i_0 , \dots, i_{r-1}$, a pivoting step (interchange of two columns)
is performed, if required, to ensure the pivot entry in row $i_j$ and
column $j$ is nonzero, and then entries to the right of the pivot are
eliminated. By recording the column swaps separately in a permutation
matrix $P$, and the eliminations in a unit upper triangular matrix $U$,
we arrive at a structured transformation of the input matrix to column
echelon form $AP^T U$.

\begin{examp}\label{exam:ech}
The following shows the structured transformation of a $7 \times 5$
matrix $A$ with row rank profile $[0, 3, 4]$ to column echelon form $C$.
$$
 \overset{AP^T}{
\begin{bmatrix}
     *&*&*&*&*\\
     *&*&*&*&*\\
     *&*&*&*&*\\
     *&*&*&*&*\\
     *&*&*&*&*\\
     *&*&*&*&*\\
     *&*&*&*&*\\
   \end{bmatrix}}
 \overset{U}{\begin{bmatrix}
    1&*&*&*&*\\
    &1&*&*&*\\
    &&1&*&*\\
    &&&1\\
    &&&&1\\
   \end{bmatrix}}
=
 \overset{C}{\begin{bmatrix}
     \overline{*}_1&&&&\\
     *&&&&\\
     *&&&&\\
     *&\overline{*}_2&&&\\
     *&*&\overline{*}_3&&\\
     *&*&*&\phantom{*}&\phantom{*}\\
     *&*&*&&\\
   \end{bmatrix}}
$$  
\end{examp}
Once an echelon form $C$ has been obtained, post-multiplication by
a diagonal matrix $D$ can be used to make the pivot elements in the
echelon form equal to~$1$, and a further post-multiplication by a unit
lower triangular $L$ can be used to eliminate entries before each pivot,
giving the canonical reduced column echelon form $R$ of $A$.

\begin{examp} \label{ex:tur}
The following shows the structured transformation of the matrix from
Example~\ref{exam:ech} to reduced column echelon form.
$$
 \overset{AP^T}{\begin{bmatrix}
     *&*&*&*&*\\
     *&*&*&*&*\\
     *&*&*&*&*\\
     *&*&*&*&*\\
     *&*&*&*&*\\
     *&*&*&*&*\\
     *&*&*&*&*\\
   \end{bmatrix}}
  \overset{U}{\begin{bmatrix}
    1&*&*&*&*\\
    &1&*&*&*\\
    &&1&*&*\\
    &&&1\\
    &&&&1\\
   \end{bmatrix}}
\overset{D}{\begin{bmatrix}
    \overline{*}_1^{-1}\\
    &\overline{*}_2^{-1}\\
    &&\overline{*}_3^{-1}\\
    &&&1\\
    &&&&1
 \end{bmatrix}}
 \overset{L}{\begin{bmatrix}
     1&&&\\
     *&1&&\\
     *&*&1&\\
     &&&1\\
     &&& &1\\
     &&& & &1\\
\end{bmatrix}}
=
 \overset{R}{\begin{bmatrix}
     1&&&&\\
     *&&&&\\
     *&&&&\\
      &1&&&\\
     &&1&&\\
     *&*&*&\phantom{*}&\phantom{*}\\
     *&*&*&&\\
   \end{bmatrix}}
$$  
\end{examp}

\begin{prop}
\label{prop:echredech}
Let $A$ be an $m \times n$ matrix with row rank profile $[i_0,\dots ,
i_{r-1}]$. Corresponding to any $n \times n$ permutation matrix $P$ such
that the submatrix of $AP^T$ comprised of rows $i_0 \dots,i_{r-1}$
has generic rank profile, there exists
\begin{itemize}
\item a unique $n \times n$ unit upper triangular matrix 
$$U=  \begin{bmatrix}
  U_1&U_2\\
  &I_{n-r}
  \end{bmatrix}
$$
such that $C = AP^T U$ is a column echelon form of $A$, and \item a unique
diagonal matrix $D = \text {Diag}(D_1 ,I_{n-r} )$ and $n \times n$ 
unit lower triangular
$$L=
  \begin{bmatrix}
    L_1\\
    & I_{n-r}
\end{bmatrix}
$$
\end{itemize}
such that $R = AP^T U DL$ is the reduced column echelon form of A.
\end{prop}

The literature contains a number of well known matrix decompositions
that reveal the row rank profile: the  common ingredient is
a permutation matrix $P$ that satisfies the requirements of
Proposition~\ref{prop:echredech}.  The Turing decomposition of the
transpose of $A$ is as shown in Example~\ref{ex:tur} except with
the matrices $U$, $D$, $L$, and $P$ inverted and appearing on the
right-hand side of the equation, and was introduced by \citet{CorJef97}
as a generalization to the rectangular case of the square decomposition
$A=LDU$ given by \citeauthor{Tur48} in his seminal \citeyear{Tur48}
paper.  The \lsp and \lqup decompositions are due to 
\citet{IbMoHu82}.  A more compact variant of \lqup is
\qlup, and the \cupd decomposition is another variation of the
\lsp decomposition that we introduce here.  The \lqup and \qlup
decompositions also involve a permutation matrix $Q$ such that the
first $r$ rows of $QA$ are equal to rows $i_0, \dots, i_{r-1}$ of $A$.
Once the permutations $P$ and $Q$ satisfying these requirements are
fixed, these five decompositions are uniquely defined and in a one-to-one
correspondence.  The following proposition links these decompositions
by defining each of them in terms of the matrices $U , C, D, L$, and $R$
of Proposition~\ref{prop:echredech}.  For completeness, the proposition
begins by recalling the definitions of the classic transformations of $A$
to column echelon form and to reduced column echelon form.

\begin{prop}
\label{prop:rprevealdecomp}  
Corresponding to an $n \times n$ permutation matrix $P$ such that rows
$i_0,\dots , i_{r-1}$ of $AP^T $ have generic rank profile, let
$U,C,D,L$ and $R$ be the matrices defined in Proposition~\ref{prop:echredech}.
The following transformations exist and are uniquely defined based
on the choice of $P$.
\begin{itemize}
\item \texttt{ColEchTrans}: $(P,U,C)$ such that $AP^TU=C$.\\
  Example:
$$
 \overset{AP^T}{
\begin{bmatrix}
     *&*&*&*&*\\
     *&*&*&*&*\\
     *&*&*&*&*\\
     *&*&*&*&*\\
     *&*&*&*&*\\
     *&*&*&*&*\\
     *&*&*&*&*\\
   \end{bmatrix}}
 \overset{U}{\begin{bmatrix}
    1&*&*&*&*\\
    &1&*&*&*\\
    &&1&*&*\\
    &&&1\\
    &&&&1\\
   \end{bmatrix}}
=
 \overset{C}{\begin{bmatrix}
     \overline{*}_1&&&&\\
     *&&&&\\
     *&&&&\\
     *&\overline{*}_2&&&\\
     *&*&\overline{*}_3&&\\
     *&*&*&\phantom{*}&\phantom{*}\\
     *&*&*&&\\
   \end{bmatrix}}
$$
\item \texttt{RedColEchTrans}: $(P, X , R)$ such that $AP^T X = R$
with $X = U DL$.\\
Example:
$$
 \overset{AP^T}{
\begin{bmatrix}
     *&*&*&*&*\\
     *&*&*&*&*\\
     *&*&*&*&*\\
     *&*&*&*&*\\
     *&*&*&*&*\\
     *&*&*&*&*\\
     *&*&*&*&*\\
   \end{bmatrix}}
 \overset{X}{\begin{bmatrix}
     *&*&*&*&*\\
     *&*&*&*&*\\
     *&*&*&*&*\\
    &&&1\\
    &&&&1\\
   \end{bmatrix}}
=
 \overset{R}{\begin{bmatrix}
     1&&&&\\
     *&&&&\\
     *&&&&\\
     &1&&&\\
     &&1&&\\
     *&*&*&\phantom{*}&\phantom{*}\\
     *&*&*&&\\
   \end{bmatrix}}
$$
\end{itemize}
Now let $\overline{L} = L^{-1} , \overline{D} = D^{-1}$ and 
$\overline{U} = U^{-1}$.
The following decompositions exist and are uniquely defined based on
the choice of $P$.
\begin{itemize}
\item \texttt{TuringDecomposition:} $(R, \overline{L}, \overline{D}, 
\overline{U} , P )$ such that $AP^T =
R\overline{L}\overline{D}\overline{U}$.\\
Example:
$$
 \overset{AP^T}{\begin{bmatrix}
     *&*&*&*&*\\
     *&*&*&*&*\\
     *&*&*&*&*\\
     *&*&*&*&*\\
     *&*&*&*&*\\
     *&*&*&*&*\\
     *&*&*&*&*\\
   \end{bmatrix}}
=
 \overset{R}{\begin{bmatrix}
     1&&&&\\
     *&&&&\\
     *&&&&\\
      &1&&&\\
     &&1&&\\
     *&*&*&\phantom{*}&\phantom{*}\\
     *&*&*&&\\
   \end{bmatrix}}
 \overset{\overline{L}}{\left[\begin{array}{ccccc}
     1&&&\\
     *&1&&\\
     *&*&1&\\
     &&&1\\
     &&& &1\\
\end{array}\right]}
\overset{\overline{D}}{\begin{bmatrix}
    \overline{*}_1\\
    &\overline{*}_2\\
    &&\overline{*}_3\\
    &&&1\\
    &&&&1
 \end{bmatrix}}
  \overset{\overline{U}}{\left[\begin{array}{ccccc}
    1&*&*&*&*\\
    &1&*&*&*\\
    &&1&*&*\\
    &&&1\\
    &&&&1\\
\end{array}\right]}
$$
\item \texttt{CUP}: $(C,\overline{U},P)$ such that $AP^T=C{\overline U}$.\\
Example:
$$
 \overset{AP^T}{
\begin{bmatrix}
     *&*&*&*&*\\
     *&*&*&*&*\\
     *&*&*&*&*\\
     *&*&*&*&*\\
     *&*&*&*&*\\
     *&*&*&*&*\\
     *&*&*&*&*\\
   \end{bmatrix}}
=
 \overset{C}{\begin{bmatrix}
     \overline{*}_1&&&&\\
     *&&&&\\
     *&&&&\\
     *&\overline{*}_2&&&\\
     *&*&\overline{*}_3&&\\
     *&*&*&\phantom{*}&\phantom{*}\\
     *&*&*&&\\
   \end{bmatrix}}
 \overset{\overline{U}}{\begin{bmatrix}
    1&*&*&*&*\\
    &1&*&*&*\\
    &&1&*&*\\
    &&&1\\
    &&&&1\\
   \end{bmatrix}}
$$

\item \texttt{LSP:} $(L',S,P)$ such that
  $A=L'SP$ with 
  \begin{itemize}
  \item $L'$ an $m \times m$ unit lower triangular with columns
  $i_0,\dots, i_{r-1}$ equal to columns
   $0,\dots, r-1$ of $CD$, and other columns those of $I_m$.
  \item $S$ an $m \times n$ semi upper triangular matrix
      with rows $i_0,\dots,i_{r-1}$
 equal to rows $0,\dots,r-1$ of $\overline{D}{\overline U}$,
and other rows zero.
\end{itemize}
Example:
$$
 \overset{AP^T}{
\begin{bmatrix}
     *&*&*&*&*\\
     *&*&*&*&*\\
     *&*&*&*&*\\
     *&*&*&*&*\\
     *&*&*&*&*\\
     *&*&*&*&*\\
     *&*&*&*&*\\
   \end{bmatrix}}
=
 \overset{L'}{\begin{bmatrix}
     1&&&\\
     *&1&&\\
     *& &1&\\
     *& & &1\\
     *& & &*&1\\
     *& & &*&*&1\\
     *& & &*&*& &1\\
\end{bmatrix}}
 \overset{S}{\begin{bmatrix}
    \overline{*}_1&*&*&*&*\\
     & \\
    & \\
    &\overline{*}_2&*&*&*\\
    &&\overline{*}_3&*&*\\
    &&&\phantom{*}\\
    &&&&\phantom{*}\\
   \end{bmatrix}}
$$
\end{itemize}
In addition to $P$, let $Q$ be an $m \times m$ permutation matrix such
that rows $0,\dots,r-1$ of $QA$ are rows $i_0, \dots, i_{r-1}$ of $A$.
Then the following decompositions exist and are uniquely defined based
on the choice of $P$ and $Q$.
\begin{itemize}
\item \texttt{LQUP:}  $(L',Q,U',P)$ such that
$A=L'QU'P$, with $L'$ the same matrix as in the \lsp
decomposition, and $U'$ an $m \times n$ matrix with 
first $r$ rows equal to the first $r$ rows of $\overline{D}\overline{U}$,
and last $n-r$ rows zero.\\
Example:
$$
 \overset{AP^T}{
\begin{bmatrix}
     *&*&*&*&*\\
     *&*&*&*&*\\
     *&*&*&*&*\\
     *&*&*&*&*\\
     *&*&*&*&*\\
     *&*&*&*&*\\
     *&*&*&*&*\\
   \end{bmatrix}}
=
 \overset{L'}{\begin{bmatrix}
     1&&&\\
     *&1&&\\
     *& &1&\\
     *& & &1\\
     *& & &*&1\\
     *& & &*&*&1\\
     *& & &*&*& &1\\
\end{bmatrix}}
Q
 \overset{U'}{\begin{bmatrix}
    \overline{*}_1&*&*&*&*\\
    &\overline{*}_2&*&*&*\\
    &&\overline{*}_3&*&*\\
    &&&\phantom{*}\\
    &&&&\phantom{*}\\
    &&&&\phantom{*}\\
   \end{bmatrix}} 
$$

\item \texttt{QLUP}: $(Q,L'',U'',P)$ with $L''$ an
  $m\times r$  unit lower triangular matrix equal to the first 
  $r$ columns of $Q^TCD$, and
  $U''$ an $r\times n$ upper triangular matrix equal to the first $r$
  rows of $\overline{D}\overline{U}$.\\
      Example:
$$
 \overset{AP^T}{
\begin{bmatrix}
     *&*&*&*&*\\
     *&*&*&*&*\\
     *&*&*&*&*\\
     *&*&*&*&*\\
     *&*&*&*&*\\
     *&*&*&*&*\\
     *&*&*&*&*\\
   \end{bmatrix}}
=
Q
 \overset{\overline{L}}{\begin{bmatrix}
     1&&&\\
     *&1&\\
     *&*&1\\
     *& & \\
     *& & \\
     *&*&*\\
     *&*&*\\
\end{bmatrix}}
 \overset{\overline{U}}{\begin{bmatrix}
    \overline{*}_1&*&*&*&*\\
    &\overline{*}_2&*&*&*\\
    &&\overline{*}_3&*&*\\
   \end{bmatrix}}
$$
\end{itemize}
\end{prop}
Note that the \lsp, \lqup, and \qlup decompositions can be transformed
from one to the other without any field operations.

\section{The \cupd matrix decomposition: algorithm and reductions}
\label{sec:algo}
In this section we propose an algorithm for the \cupd decomposition, and
show how each of the other four decompositions and two transformations
of Proposition~\ref{prop:rprevealdecomp} can be recovered via the \cupd
decomposition.  For clarity we postpone to Section~\ref{sec:discussion}
the discussion of the advantages, in terms of time and space complexities,
of the \cupd decomposition algorithm over other Gaussian elimination
based agorithms in the literature.

All algorithms presented in this section are recursive and operate
on blocks; this groups arithmetic operations into large matrix
multiplication updates of the form  $C =\alpha A B + \beta C$.
Reducing dense linear algebra to matrix multiplication
not only leads to a reduced asymptotic time
complexity because of the subcubic exponent of matrix multiplication,
but also ensures good efficiency in practice thanks to the availability
of highly optimized subroutines \citep[see for
example][]{DumasGautierPernet}.

\subsection{Space-sharing storage and in-place computations}
\label{sec:inplace}
Dealing with space complexity, we make the assumption that a field element as
well as indices use an atomic memory space.
We distinguish two ways to reduce the memory consumption of our algorithms:

\paragraph*{Space-sharing storage}
All the algorithms described here, except for matrix multiplication,
store their matrix outputs (excluding permutations) in the space
allocated to their inputs.  For example, when solving the triangular
system with matrix right-hand side $UX=B$, the input matrix $B$
will be overwritten by the output matrix $X=U^{-1}B$.  In the case
of matrix decompositions, the output consists of one or two
permutations that can be stored using an amount of space linear in
the sum of the matrix dimensions, and two structured matrices that
can be stored together within the space of the input matrix. We
call this a space-sharing storage. 

The space-sharing storage for the \cupd and \texttt{ColEchTrans}
of a matrix of rank $r$ stores the first $r$ columns of the lower
triangular factor below the first $r$ rows of the upper triangular
factor in the same matrix.   Overlap is avoided by storing only the
nontrivial diagonal entries.  Here is an example for a \cupd
decomposition.
$$
 \overset{C}{\begin{bmatrix}
     \overline{*}_1&&&&\\
     *&&&&\\
     *&&&&\\
     *&\overline{*}_2&&&\\
     *&*&\overline{*}_3&&\\
     *&*&*&\phantom{*}&\phantom{*}\\
     *&*&*&&\\
   \end{bmatrix}}
 \overset{\overline{U}}{\begin{bmatrix}
    1&*&*&*&*\\
    &1&*&*&*\\
    &&1&*&*\\
    &&&1\\
    &&&&1\\
   \end{bmatrix}} \longrightarrow
 \begin{bmatrix}
     \overline{*}_1&*&*&*&*\\
     *&&*&*&*\\
     *&&&*&*\\
     *&\overline{*}_2&&&\\
     *&*&\overline{*}_3&&\\
     *&*&*&\phantom{*}&\phantom{*}\\
     *&*&*&&\\
   \end{bmatrix}
$$

The \texttt{RedColEchelonTrans}
can be stored in a space-sharing manner up to some
permutation.  Indeed, any transformation to reduced column echelon
form of rank $r$ can be written as
  $$AP^T { \begin{bmatrix}
    X_1&X_2\\&I_{n-r}
  \end{bmatrix}} = R = Q^T  { \begin{bmatrix}
    I_r&\\
    \underline{R}&0
  \end{bmatrix}}$$ for some permutation matrix $Q$.
Up to this permutation of rows, this allows the space-sharing storage
$${ \begin{bmatrix}
 X_1&X_2\\ \underline{R}
\end{bmatrix}}.$$

Similar space-sharing storage formats are possible for the \qlup,
\lsp, \lqup and \texttt{TuringDecomposition}.



\paragraph*{In-place computation}
More importantly, we will also focus on the intermediate memory
allocations of the algorithms presented. As the algorithms rely heavily
on matrix multiplication, one needs to take into account the memory
allocations in the matrix multiplication algorithm.  Using the classical
$O(n^3)$ algorithm, one can perform the operation $C \leftarrow \alpha A B
+ \beta C$ with no additional memory space beyond that required to store
the input and output, but this is no longer the case with Strassen and
Winograd's $O(n^{2.81})$ algorithms \citep[for a detailed analysis see
for example][]{Huss-Lederman:1996:mai,BoyerDumasPernetZhou:2009:Winograd}.
Consequently, we will consider matrix multiplication as a black-box
operation, and analyze the memory complexity of our algorithms
independently of it.  This leads us to introduce the following definition.

\begin{defn} \label{def:inplace}
A linear algebra algorithm is called {\em in-place} if it does not
involve more than $O(1)$ extra memory allocations for field elements
for any of its operations except possibly in the course of matrix
multiplications.
\end{defn}
In particular, when {\em classical} matrix multiplication is used,
in-place linear algebra algorithms only require a constant amount of
extra memory allocation.

\subsection{Basic building blocks: \texttt{MM}, \texttt{TRSM}, \texttt{TRMM},
   \texttt{TRTRI}, \texttt{TRULM},  \texttt{TRLUM}} \label{subsec:routines}
We assume that Algorithm~\ref{alg:mm} for matrix 
multiplication is available.
\begin{algorithm} 
\KwData{$A$ an $m\times \ell$ matrix.}
\KwData{$B$ an $\ell \times n$ matrix.}
\KwData{$C$ an $m \times n$ matrix that does not overlap 
  with either $A$ or $B$.}
\KwData{$\alpha$ a scalar.}
\KwData{ $\beta$ a scalar.}
\KwEnsures{ $C\leftarrow \alpha A B + \beta C$.}
\caption{\mm\!\!($A$, $B$, $C$, $\alpha$, $\beta$)}
\label{alg:mm}
\end{algorithm}
We also use the well-known routines \trsm and \trtri, defined
in the level~3 BLAS legacy \citep{Dongara:1990:BLAS3} and the LAPACK
library \citep{LAPACKUsersGuide}.  The \trsm routine simultaneously
computes several linear system solutions $X$ from an invertible triangular
matrix $A$ and a matrix right-hand side $B$.  The matrix $A$ can be
lower or upper triangular, unit or non-unit triangular, left-looking
($AX = B$)  or right-looking ($XA = B$).  Algorithm \ref{alg:trsm} below
illustrates the case ``right-looking, upper, non-unit,'' and shows how
to incorporate matrix multiplication; the algorithm is clearly 
in-place \citep[\S4.1]{DuGiPe08} and has running time $O(\max\{m,n\}\,
n^{\omega-1})$.
\begin{algorithm}
\DontPrintSemicolon
\KwData{$U$ an $n\times n$ invertible upper triangular matrix.}
\KwData{$B$ an  $m\times n$ matrix.} 
\KwEnsures{$B\leftarrow  B U^{-1}$.}
\Begin{
    \eIf{$n=1$}{
      \lFor{$i\leftarrow 0,\dots, m-1$}{$b_{i,0}\leftarrow b_{i,0} / u_{0,0}$\;}
    } {
      $k\leftarrow \lfloor \frac{n}{2}\rfloor$\;
      \tcc{\footnotesize  $B= \begin{bmatrix}  B_1&B_2 \end{bmatrix}$ with  $B_1$ $m\times k$
              and $U= \begin{bmatrix} U_1&V\\&U_2 \end{bmatrix}$ with $U_1$ $k\times k$
            }
      \trsm\!\!$(\text{Right, Up, NonUnit},\submatrix{U}{0}{k}{0}{k}, \submatrix{B}{0}{m}{0}{k})$
      \tcc*{\small $B_1 \leftarrow  B_1U_1^{-1}$}
      \mm\!\!$(\submatrix{B}{0}{m}{0}{k}, \submatrix{U}{0}{k}{k}{n}, \submatrix{B}{0}{m}{k}{n},-1,1)$
      \tcc*{\small $B_2\leftarrow B_2-B_1V$}
      \trsm\!\!$(\text{Right, Up, NonUnit},\submatrix{U}{k}{n}{k}{n}, \submatrix{B}{0}{m}{k}{n})$
      \tcc*{\small $B_2\leftarrow B_2U_2^{-1}$}
    }
  }
\caption{\trsm\!\!(Right, Up, NonUnit, $U$,$B$)}
\label{alg:trsm}
\end{algorithm}

We remark that an algorithm similar to Algorithm~\ref{alg:trsm} can be
written for the so-called \trmm routine, which computes the product of
a triangular matrix by an arbitrary matrix, in the same eight variants
\citep[for the square case see for example][\S6.2.1]{DuGiPe08}.
Clearly, \trmm has the same in-place and complexity features as \trsm.

The \trtri routine inverts a triangular matrix that can be either upper or
lower triangular, unit or non-unit triangular.  Algorithm \ref{alg:trtri}
illustrates the case ``upper, non-unit'' and shares the following features
with the three other variants: it reduces to matrix multiplication via
two calls to \trsm in half the dimension, has a cost in $O(n^\omega)$,
and is in-place.
\begin{algorithm}
  \DontPrintSemicolon
\KwData{$U$ an $n\times n$ invertible upper triangular matrix.}
\KwEnsures{ $U \leftarrow U^{-1}$.}
\Begin{
    \eIf{$n=1$}    {$u_{0,0}\leftarrow 1/u_{0,0}$\;}
        {
          $ k\leftarrow \lfloor \frac{n}{2}\rfloor$\;
          \tcc{\footnotesize $U= \begin{bmatrix} U_1&V\\&U_2 \end{bmatrix}$ with $U_1$ $k \times k$, s.t. $U^{-1} = \begin{bmatrix} U_1^{-1} & -U_1^{-1}VU_2^{-1} \\&U_2^{-1} \end{bmatrix}$}
          $\trsm(\text{Right, Up, NonUnit}, \submatrix{U}{k}{n}{k}{ n},  \submatrix{U}{0}{k}{k}{ n})$
          \tcc*{\small $V \leftarrow VU_2^{-1}$}
          $\trsm(\text{Left, Up, NonUnit}, \submatrix{U}{0}{k}{0}{ k}, \submatrix{U}{0}{k}{k}{n})$
          \tcc*{\small $V \leftarrow U_1^{-1}V$}
          $\submatrix{U}{0}{k}{k}{ n} \leftarrow - \submatrix{U}{0}{k}{k}{n}$
          \tcc*{$V\leftarrow -V$}
          $\trtri(\submatrix{U}{0}{k}{0}{k})$
          \tcc*{\small $U_1\leftarrow U_1^{-1}$}
          $\trtri(\submatrix{U}{k}{n}{k}{n})$
          \tcc*{\small $U_2 \leftarrow U_2^{-1}$}
          }
  }
  \caption{\trtri\!\!(Up, NonUnit, $U$)}
  \label{alg:trtri}
\end{algorithm}

We conclude this section by introducing the \trulm and \trlum
routines: given a unit lower triangular matrix $L$ and an upper
triangular matrix $U$ stored one next to the other within the same
square matrix, they return the products $UL$ and  $LU$,  respectively.
Unlike \trsm and \trtri, these routines are neither  BLAS nor  LAPACK
routines, and to our knowledge have not yet been described elsewhere.
Algorithms~\ref{alg:trulm} and~\ref{alg:trlum} show how they can be
implemented in-place and at cost $O(n^\omega)$.  The routine~\trulm is the
key step that  enables us to compute the reduced echelon form in-place and
in subcubic time (see Algorithm~\ref{alg:redcolechelon}), while \trlum
will be used to derive a fast and in-place method for matrix products
of the form $B\leftarrow A\times B$ (see Algorithm~\ref{alg:inplacemm}).

\begin{algorithm}
  \DontPrintSemicolon

\KwData{$A = [L\backslash U]$ an $n\times n$ matrix.}
\KwEnsures{ $A \leftarrow UL$.}
\Begin{
    \If{$n>1$}{
      $ k\leftarrow \lfloor \frac{n}{2}\rfloor$\;
      \tcc{\footnotesize $A=\begin{bmatrix}L_1\backslash U_1 &
        U_2\\ L_2&L_3\backslash U_3\end{bmatrix}$ and
        $LU=\begin{bmatrix}X_1&X_2\\ X_3&X_4\end{bmatrix}$ with $L_1, U_1, X_1$ $k\times k$} 
      $\trulm(\submatrix{A}{0}{k}{0}{k})$
      \tcc*{\small {$X_1\leftarrow  U_1L_1$}}
      $\mm(\submatrix{A}{0}{k}{k}{n},\submatrix{A}{k}{n}{0}{k},
      \submatrix{A}{0}{k}{0}{k}, 1,1 )$
      \tcc*{\small $X_1 \leftarrow  X_1 + U_2L_2$}
      $\trmm(\text{Right, Low, Unit}, \submatrix{A}{k}{n}{k}{n},\submatrix{A}{0}{k}{k}{n})$
      \tcc*{\small $X_2 \leftarrow  U_2L_3$}
      $\trmm(\text{Left, Up, NonUnit}, \submatrix{A}{k}{n}{k}{n},\submatrix{A}{k}{n}{0}{k})$
      \tcc*{\small $X_3 \leftarrow  U_3L_2$}
      $\trulm(\submatrix{A}{k}{n}{k}{n})$
      \tcc*{\small $X_4\leftarrow  U_3L_3$}
    }
  }
\caption{\trulm($A$)}
\label{alg:trulm}
\end{algorithm}

\begin{algorithm}
  \DontPrintSemicolon

\KwData{$A = [L\backslash U]$ an $n\times n$ matrix.}
\KwEnsures{ $A \leftarrow LU$.}
\Begin{
    \If{$n>1$}{
      $ k\leftarrow \lfloor \frac{n}{2}\rfloor$\;
      \tcc{\footnotesize $A=\begin{bmatrix}L_1\backslash U_1 &
        U_2\\ L_2&L_3\backslash U_3\end{bmatrix}$ and
        $UL=\begin{bmatrix}X_1&X_2\\ X_3&X_4\end{bmatrix}$ with $L_1, U_1, X_1$ $k\times k$} 
      $\trlum(\submatrix{A}{k}{n}{k}{n})$
      \tcc*{\small $X_4\leftarrow L_3U_3$}
      $\mm(\submatrix{A}{k}{n}{0}{k},\submatrix{A}{0}{k}{k}{n},
      \submatrix{A}{k}{n}{k}{n}, 1,1 )$
      \tcc*{\small $X_4 \leftarrow X_4 + L_2U_2$}
      $\trmm(\text{Left, Low, Unit},\submatrix{A}{0}{k}{0}{k},\submatrix{A}{0}{k}{k}{n})$
      \tcc*{\small $X_2 \leftarrow  L_1U_2$}
      $\trmm(\text{Right, Up, NonUnit}, \submatrix{A}{0}{k}{0}{k},\submatrix{A}{k}{n}{0}{k})$
      \tcc*{\small $X_3 \leftarrow  L_2U_1$}
      $\trlum(\submatrix{A}{0}{k}{0}{k})$
      \tcc*{\small {$X_1\leftarrow  L_1U_1$}}
    }
  }
\caption{\trlum($A$)}
\label{alg:trlum}
\end{algorithm}

\subsection{The \cupd matrix decomposition algorithm} \label{ssec:cupalg}

We now present Algorithm~\ref{alg:cup}, a block recursive algorithm for
factoring any $m\times n$ matrix $A$ as $A = CUP$ with $C$ a column
echelon form revealing the row rank profile of $A$, $U$ a unit upper
triangular matrix, and $P$ a permutation matrix.  It is a variation on
the \lsp and \lqup decomposition algorithms of \citet{IbMoHu82}, similar
to the ones presented by \citet{Jea06} and
\citet{DuGiPe08}.  Similar to the basic
building blocks of Section~\ref{subsec:routines}, the description using
submatrix indices shows that the entire algorithm can be performed 
in-place: each recursive call transforms a block into a block of the form
$[C\backslash U]$, and the remaining operations are a \trsm, a matrix
multiplication, and applying row and column permutations.  The \cupd
algorithm also handles the case where $m>n$. The ability to work in-place
and handle matrices of arbitrary shape are the main advantages of the
algorithm for \cupd decomposition over the \texttt{LSP} or \texttt{LQUP}
of \citet{IbMoHu82}, as will be discussed in Section~\ref{sec:discussion}.

\begin{algorithm}
  \DontPrintSemicolon
  \KwData{$A$ an $m \times n$ matrix.}
  \KwResult{$(r,[i_0,\dots,i_{r-1}],P)$.}
  \KwEnsures{$A \leftarrow [C\backslash U]$ such that, embedding $U$ in an
    $n\times n$ unit upper triangular matrix makes $A=CUP$ a \cupd decomposition of $A$, 
    and $r$ and $[i_0,\dots,i_{r-1}]$ are the rank and the row rank profile of $A$.}
  \Begin{
      \If{$m=1$}{
        \If{$A\neq 0$}{
          $i \leftarrow $ column index of the first nonzero entry of $A$\;
          $P \leftarrow T_{0,i}$, the transposition of indices $0$ and $i$\;
          $A \leftarrow A P$\;
          \lFor{$i\leftarrow 2\dots n$}{$A_{0,i}\leftarrow A_{0,i}A_{0,0}^{-1}$\;}
          \Return{$(1, [0], P)$}
        }{
          \Return{$(0, [], I_n)$}
        }
      }
      $k \leftarrow \lfloor \frac{m}{2} \rfloor$
      \tcc*{\small $A=
        \left[
        \begin{array}{c}
          A_1\\
          \hline
          A_2
        \end{array}\right]$ with $A_1$ $k \times n$
      }
      \lnl{cup:step:rec1}$(r_1,[i_0,\dots,i_{r_1-1}],P_1) \leftarrow \cupd(\submatrix{A}{0}{k}{0}{n})$ 
         \tcc*{\small $A_1 \leftarrow C_1\begin{bmatrix} U_1 &  V_1 \end{bmatrix}P_1 $}
         \If{$r_1 = 0$}{
           $(r_2,[i_0,\dots,i_{r_2-1}],P_2)\leftarrow\cupd(\submatrix{A}{k}{m}{0}{n})$\;
           \Return{$(r_2,[i_0+k,\dots,i_{r_2-1}+k],P_2)$}  
         } 
         $\submatrix{A}{k}{m}{0}{n} \leftarrow
         \submatrix{A}{k}{m}{0}{n}P_1^T $
\tcc*{
$  \begin{bmatrix}\tilde{A}_{21}&\tilde{A}_{22} \end{bmatrix} \leftarrow \begin{bmatrix} A_{21}&A_{22} \end{bmatrix}P_1^T$
}
         \tcc{\small $A=
           \begin{bmatrix}
             C_1\backslash 
             \begin{array}{c}
               U_1\\0
             \end{array}
&             \begin{array}{c}
               V_1\\0
             \end{array}\\
             \tilde A_{21}      & \tilde A_{22}
           \end{bmatrix}
           $ with $C_1\backslash U_1$  $k\times r_1$ and $V_1$  $r_1\times(n-r_1)$}
         \lnl{cup:step:trsm} $\trsm(\text{Right},\text{Upper},\text{Unit},\submatrix{A}{0}{r_1}{0}{r_1},\submatrix{A}{k}{m}{0}{r_1}$)
         \tcc*{\small $G \leftarrow A_{21}U_1^{-1}$}
         \If{$r_1=n$}{
            \Return {$(r_1,[i_0,\dots,i_{r_1-1}],P_1)$}
         }
         \lnl{cup:step:mm}$\mm(\submatrix{A}{k}{m}{0}{r_1} , \submatrix{A}{r_1}{n}{0}{r_1},\submatrix{A}{k}{m}{r_1}{n},-1,1)$
	 \tcc*{\small $H \leftarrow A_{22} - GV_1$}
	 \lnl{cup:step:rec2}$(r_2,[j_0,\dots,j_{r_2-1}],P_2) \leftarrow
         \cupd(\submatrix{A}{k}{m}{r_1}{n})$ 
         \tcc*{\small $H\leftarrow
            C_2 U_2  P_2$}
         $\submatrix{A}{0}{r_1}{r_1}{n} \leftarrow  \submatrix{A}{0}{r_1}{r_1}{n} P_2^T $
         \tcc*{\small $\tilde{V}_1\leftarrow V_1P_2^T$}

         \lnl{cup:step:permute}
         \For{$j\leftarrow r_1\dots r_1+r_2$}{
           
           $\rowofmatrix{A}{j}{n}{j} \leftarrow \rowofmatrix{A}{j}{n}{j+k-r_1}$
           \tcc*{\small Moving $U_2, V_2$ up next to $V_1$}
           $ \colofmatrix{A}{j}{m}{j+k-r_1} \leftarrow [0]$\;
         }
         $P \leftarrow \text{Diag}(I_{r_1}, P_2) P_1$\;
         \Return {$ (r_1+r_2,[i_0,\dots,i_{r_1-1},j_0+k,\dots,j_{r_2-1}+k], P)$}
    }
    \caption{\cupd\!\!($A$)}
    \label{alg:cup}
  \end{algorithm}
\begin{figure}[htbp]
\begin{center}
 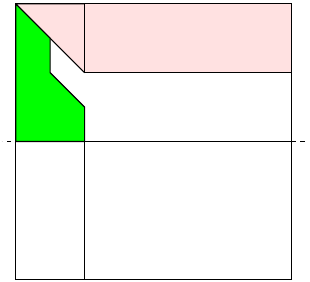
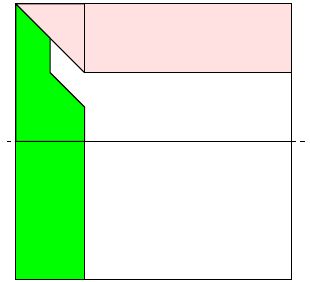
 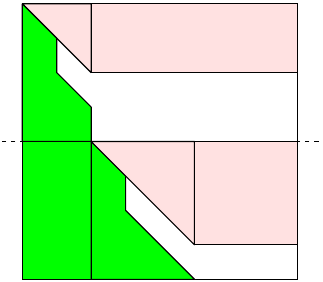
 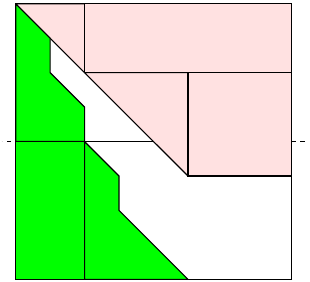
\end{center}
\caption{Illustration of Algorithm~\ref{alg:cup}: steps \ref{cup:step:rec1},
  \ref{cup:step:trsm}, \ref{cup:step:rec2}, and \ref{cup:step:permute}.}
\label{fig:cup}
\end{figure}

Figure~\ref{fig:cup} shows the state of the input matrix after the
four main steps (steps \ref{cup:step:rec1}, \ref{cup:step:trsm},
\ref{cup:step:rec2}, and \ref{cup:step:permute}).  At
step~\ref{cup:step:permute}, the zero rows between blocks $V_1$ and
$[U_2 \, V_2]$ are shifted down.

\begin{prop}\label{prop:cup}
  Algorithm~\ref{alg:cup} computes a \cupd decomposition of the input matrix~$A$
  together with its rank $r$ and row rank profile $[i_0,\dots, i_{r-1}]$.
\end{prop}

\begin{pf}
  It is clear from the description of the algorithm that $P$ is a permutation matrix, 
  $U$ is unit upper triangular, and $C$ is in column echelon form.
  To check that $A = CUP$, one can proceed exactly as \citet[\S3.1]{Jea06}. 
  \qed
\end{pf}

\subsection{\colechelon transormation via a \cupd decomposition}
\label{ssec:cuptoech}
Proposition~\ref{prop:rowech} states how to compute 
the required transformation matrix $X$ such that 
$AX=C$ is in echelon form. 
\begin{prop}
\label{prop:rowech}
  Let $A=CUP$ be the \cupd decomposition of an $m\times n$ matrix $A$
  of rank $r$ computed with Algorithm \ref{alg:cup}. 
  Let $X$ be the inverse of $U$ embedded in an $n\times n$ unit upper triangular
  matrix.
  Then $AP^TX=E$ is a transformation of $A$ to column echelon form.
\end{prop}
\begin{pf}
One need only verify that   $CX^{-1}=CUP=A$.
\qed
\end{pf}
  \begin{algorithm}
   \DontPrintSemicolon
   \KwData{$A$ an $m \times n$ matrix over a field.}
   \KwResult{$(r,[i_0,\dots,i_{r-1}],P)$ where $P$ is a permutation matrix.}
   \KwEnsures{$A\leftarrow[C\backslash X]$ such that, embedding $X$ in an
     $n\times n$ unit upper triangular matrix makes $AP^TX=C$ a transformation of $A$ to column echelon form,
     $r=\text{rank}(A)$ and $[i_0,\dots,i_{r-1}]$ is the row rank profile of $A$.}
  \Begin{
        $(r,[i_0,\dots,i_{r-1}],P)\leftarrow\cupd(A)$\;
      \tcc{$A=
        \begin{bmatrix}
          C \backslash U_1 & 
          \begin{array}{c}
            U_2\\0
          \end{array}
        \end{bmatrix}$ where $U_1$ is $r\times r$ upper triangular
      }
      $\trsm(\text{Left},\text{Upper},\text{Unit},\submatrix{A}{0}{r}{0}{r},\submatrix{A}{0}{r}{r}{n})$ \tcc*{$M\leftarrow U_1^{-1}U_2$}
      $\submatrix{A}{0}{r}{r}{n}\leftarrow -\submatrix{A}{0}{r}{r}{n}$ \tcc*{$N\leftarrow -M$}
      $\trtri(\text{Upper},\submatrix{A}{0}{r}{0}{r})$  \tcc*{$U\leftarrow U^{-1}$}
      \KwRet{$(r,[i_0,\dots,i_{r-1}],P)$}
    }
    \caption{$\colechelon (A)$}\label{alg:colechelon}
  \end{algorithm}
Algorithm~\ref{alg:colechelon} shows how the computation of the column
echelon form can be done in-place and reduces to that of \cupd, \trsm
and \trtri and therefore has a time complexity of \GO{mnr^{\omega-2}}.

\subsection{\redcolechelon transformation via \cupd decomposition}
\label{ssec:cuptoredech}

Proposition~\ref{prop:redech} states how the reduced column
echelon form can be computed from the \cupd decomposition.
Recall that $T_{i,j}$ denotes the permutation
matrix that swaps indices $i$ and $j$ and leaves the other elements 
unchanged.
\begin{prop}
\label{prop:redech}  Let $(r,[i_0,\dots,i_{r-1}],P)$ and $[C\backslash Y]$ be the output of
Algorithm~\ref{alg:colechelon} called on an $m\times n$ matrix $A$. As
previously, assume that the matrix $Y$ is embedded in an $n\times n$ unit upper
triangular matrix. 
Let $Q= T_{0,i_0}T_{1,i_1}\dots T_{r-1,i_{r-1}}$ so that $QC=
\begin{bmatrix}
  L_1\\L_2
\end{bmatrix}
$, where $L_1$ is an $r\times r$ nonsingular lower triangular matrix.
Let $$
R=
C
\begin{bmatrix}
L_1^{-1}
\\&I_{m-r}  
\end{bmatrix} 
\text{ and } 
X= 
Y
\begin{bmatrix}
  L_1^{-1}\\&I_{m-r}
\end{bmatrix}
.$$
Then $AP^TX=R$ is a transformation of $A$ to reduced column echelon form.
\end{prop}

\begin{pf}
We start by showing that $R$ is in reduced column echelon form.
Recall that the permutation matrix $Q$ is such that row $k$ of a matrix
$A$ is row $i_k$ of $Q^TA$.
Hence for $j\in\{0\dots r-1\}$, row $i_j$ of $R$ is row
$j$ of the identity matrix $I_m$.

Consider column $j$ of $R$. From the above remark, it has a $1$ in row
$i_{j}$. We now show that any coefficient above it is zero.
Let $i<i_{j}$, then  $R_{i,j}=\sum_{k=0}^mC_{i,k}[L^{-1}]_{k,j} $.
As $C$ is in column echelon form and $i<i_{j}$, $C_{k,i}=0$ for any $k\geq j$. Since $L^{-1}$ is
lower triangular, $L^{-1}_{k,j}=0$ for any $k<j$, hence $R_{i,j}=0$.

It remains to verify 
$$
  AX= CUPP^TY \begin{bmatrix}  L_1^{-1}\\&I_{m-r}\end{bmatrix} 
       = R.
$$
\qed
\end{pf}
  \begin{algorithm}
  \DontPrintSemicolon
   \KwData{$A$ an $m \times n$ matrix over a field.}
   \KwResult{$(r,[i_0,\dots,i_{r-1}],P)$ where $P$ is a permutation matrix.}
   \KwEnsures{$A\leftarrow
     \begin{bmatrix}
       X_1&X_2\\
       R_2&0
     \end{bmatrix} $ such that
     if $X=\begin{bmatrix}X_1&X_2\\&I_{n-r}\end{bmatrix}$
     and $Q=T_{0,i_0}T_{1,i_1}\dots T_{r-1,i_{r-1}}$ and  
     $R=Q^T\begin{bmatrix} I_r&\\R_2&0 \end{bmatrix}$ then 
     $AP^TX=R$ is a transformation of $A$ to reduced column echelon form,
     $r=\text{rank}(A)$, and $[i_0,\dots,i_{r-1}]$ is the column rank profile of $A$. }
  \Begin{
        $(r,[i_0,\dots,i_{r-1}],P)\leftarrow \colechelon(A)$\;
      \tcc{Notations: $A=
        \begin{bmatrix}
          C\backslash \begin{array}{cc}
            T_1&T_2
          \end{array}
        \end{bmatrix}$ and $X_2\leftarrow T_2$.
      }
      \For{$j\leftarrow 0\dots r-1$}{
        Swap(\rowofmatrix{A}{j}{0}{j},\rowofmatrix{A}{i_j}{0}{j})
        \tcc*{$
      \begin{bmatrix}
        L_1\\L_2
      \end{bmatrix}
      \leftarrow QC$}
      }
      $\trsm(\text{Right},\text{Lower},\text{Unit},\submatrix{A}{0}{r}{0}{r},\submatrix{A}{r}{m}{0}{r})$
      \tcc*{$R_2\leftarrow L_2L_1^{-1}$}
      $\trtri(\text{Lower},\submatrix{A}{0}{r}{0}{r})$
      \tcc*{$N\leftarrow L_1^{-1}$}
      $\trulm(\submatrix{A}{0}{r}{0}{r})$
      \tcc*{$X_1\leftarrow T_1N$}
      \KwRet{$(r,[i_0,\dots,i_{r-1}],P)$}
    }
    \caption{$\redcolechelon(A)$}\label{alg:redcolechelon}
  \end{algorithm}
Algorithm \ref{alg:redcolechelon} shows how the computation of the
reduced column echelon form transformation can be done in-place and
reduces to that of \cupd, \trsm, \trtri, and \trulm and therefore has
a time  complexity of \GO{mnr^{\omega-2}}.

\section{Discussion}
\label{sec:discussion}

Because it is the building block of dense linear algebra algorithms,
Gaussian elimination appears in many different forms in the literature
and in software implementations.  Softwares are mostly based on either
a transformation to echelon form, or one of the decompositions of
Proposition~\ref{prop:rprevealdecomp}, from which the usual computations
such as linear system solving, determinant, rank, rank profile,
nullspace basis, etc.\/ can be derived.  We discuss here the advantages
of the \cupd decomposition algorithm presented in the previous section
compared to some other algorithms in the literature.

In our comparison we only consider algorithms that reduce to matrix
multiplication. Indeed, reduction to matrix multiplication is the
only way to benefit from both the best theoretical complexity and the
practical efficiency of highly optimized implementations of the level
3 BLAS routines \citep{DuGiPe08}.

To our knowledge, the algorithms achieving subcubic time 
complexity for elimination of rank
deficient matrices are
\begin{itemize}
\item the \lsp and \lqup algorithms of \citet{IbMoHu82} for computing an \lsp
and \lqup decomposition,
\item the \gauss and \gaussjordan algorithms \citep[Algorithms
2.8 and 2.9]{Sto00}
for computing an echelon and reduced echelon form transforms, and
\item the \texttt{StepForm} algorithm \citep[][\S 16.5]{KelGeh85,BuClSh97}
for computing an echelon form.
\end{itemize}

For the sake of completeness, we recall the \texttt{GaussJordan}
and \texttt{StepForm} algorithms  in~\ref{app:storj}
and~\ref{app:step-form}, together with an analysis of their complexity.
Note that already in his seminal paper, \citet{St69} proposed an
$\GO{n^\omega}$ algorithm to compute the inverse of a matrix under
strong genericity assumptions on the matrix. The \texttt{GaussJordan}
algorithm can be viewed as a generalization: it inverts any nonsingular
matrix without assumption, and computes the reduced echelon form and a
transformation matrix in the case of  singular matrices.

Note that all references to \qlup decomposition that we know of
(namely
\citep[\S 3.4.9]{GoVa96} and \citep{Jeffrey:2010}) do not mention any
subcubic time algorithm, but as it can be derived as a slight modification
of Algorithm~\ref{alg:cup}, we will also take this variant into account
for our comparison.

Our comparison of the various algorithms will be based on both space
complexity (checking whether the algorithms are in-place) and a finer
analysis of the time complexity which considers the constant of the
leading term in the asymptotic estimate.

\subsection{Comparison of the \cupd decomposition with \lsp, \lqup, \qlup}

\paragraph*{\lsp}
As discussed in Section~\ref{sec:inplace}, the \lsp decomposition can
not enjoy a space-sharing storage, unless its $L$ matrix is compressed by
a column permutatin $Q$. Now when designing a block recursive algorithm
similar to Algorithm~\ref{alg:cup}, one needs to compress the $S$ matrix
returned by the first recursive call, in order to use a \trsm update
similar to operation~\ref{cup:step:trsm}. This requires to either
allocate an $r\times r$ temporary matrix, or to do the compression by
swapping rows of $S$ back and forth.  Instead, if the upper triangular
matrix is kept compressed during the whole algorithm, this becomes an
\lqup decomposition algorithm.

\paragraph*{\lqup}
Although \lqup decomposition can use a space sharing storage,
the intermediate steps of a block recursive algorithm derived from
Algorithm~\ref{alg:cup} would require additional column permutations
on the $L$ matrix to give it the uncompressed shape. Instead, if one
chooses to compute the \lqup decomposition with a compressed $L$ matrix,
this really corresponds to the \cupd decomposition, up to row and column
scaling by the pivots.

\paragraph*{\qlup}
The \qlup decomposition can also be computed by an adaptation of
Algorithm~\ref{alg:cup} where rows of the lower triangular matrix have
to be permuted. Such an algorithm for the \qlup decomposition would then
share the same advantages as the \cupd decomposition algorithm but the
following three reasons make the \cupd decomposition preferable.
\begin{itemize}
\item The overhead of row permutations on the lower part of the matrix
  might become costly especially with sparse matrices.
\item Part of the structure of the matrix $C$ is lost when considering
  $L,Q$ instead: in $C$, any coefficient above a pivot, in a non-pivot
  row is known to be zero by the echelon structure, whereas this same
  coefficient in $L$ has to be treated as any other coefficient, and be
  assigned the zero value.
\item It seems difficult to implement an efficient storage for the
  permutation $Q$ (as can be done for $P$, using LAPACK
  storage of permutations). One could think of setting
  $Q=T_{0,i_0}T_{1,i_1}\dots T_{r-1,i_{r-1}}$ after the algorithm has completed,
  as it is done for Algorithm~\ref{alg:redcolechelon}. However this
  permutation does not correspond to the permutation that was applied
  to the non-pivot rows of $L$ during the process of the algorithm (call
  it $\tilde Q$).  We could not find any subquadratic time algorithm to
  generate this permutation $\tilde Q$ from the two permutations $\tilde
  Q_1$ and $\tilde Q_2$ returned by the recursive calls.
\end{itemize}

The four matrix decompositions \lsp, \lqup, \pluq, and \cupd are
mathematically equivalent: they can all be computed by a dedicated
algorithm with the same amount of arithemtic operations, and any
conversion from one to another only involves permutations and pivot
scaling.  Among them, the \cupd decomposition stands out as the most
natural and appropriate one from the computational point of view: it
can use a space-sharing storage and be computed in place and with the
least amount of permutations.

\subsection{A rank sensitive time complexity}\label{sec:complexity}

As one may expect, the rank of the input matrix affects the time
complexity of the \cupd algorithm.  For example, using a naive cubic-time
algorithm for matrix multiplication the \cupd decomposition requires
$2mnr-(n+m)r^2+\frac{2}{3}r^3$ field operations for an $m\times n$
input matrix of rank $r$.

Assuming a subcubic algorithm for matrix multiplication, the
analysis in the literature for most Gaussian elimination algorithms
is not rank sensitive.  For example, the running time of the
\lsp and \lqup algorithms~\citep{IbMoHu82} is only shown to be
$\GO{m^{\omega-1}n}$, assuming $m \leq n$.  Following the analysis
of the \texttt{GaussJordan} algorithm~\citep[Algorithm~2.8]{Sto00},
we give in Proposition~\ref{prop:cup:cpxity} a rank sensitive
complexity for Algorithm~\ref{alg:cup} computing the  \cupd
decomposition of an input matrix of arbitrary shape.  According
to the reductions of Section~\ref{sec:algo}, the rank sensitive
complexity bound of Proposition~\ref{prop:cup:cpxity} also holds for
the computation of all other decompositions and transformations of
Proposition~\ref{prop:rprevealdecomp}.

\begin{prop}\label{prop:cup:cpxity}
Algorithm~\ref{alg:cup} computes a \cupd decomposition of an $m\times n$ matrix
of rank $r$ using \GO{mnr^{\omega-2}} field operations.
\end{prop}

\begin{pf}
Denote by $T_\cupd(m,n,r)$ be the number of field operations required by
Algorithm~\ref{alg:cup} for an $m\times n$ matrix $A$ of rank $r$.

In the following, we assume without loss of generality that $m$ is a power of~2.
Following \citet{Sto00} we count a comparison with zero as a field operation.
Then, when $r = 0$ (that is, $A$ is the zero matrix), we have $T(m,n,r) = \GO{mn}$.
As in the algorithm, let $r_1$ be the rank of $A_1$ and 
let $r_2$ be the rank of $H$. Then
$$
T_\cupd(m,n,r) =
\left\{
\begin{array}{ll}
\GO{n} & \text{ if } m=1,\\ 
T_\cupd(\frac{m}{2},n,0) + T_\cupd(\frac{m}{2},n,r_2) +\GO{mn}& \text{ if } r_1 =0,\\
T_\cupd(\frac{m}{2},n,n) + \GO{\frac{m}{2}n^{\omega-1}} & \text{ if } r_1 = n,\\
T_\cupd(\frac{m}{2},n,r_1) +T_\cupd(\frac{m}{2},n-r_1,r_2) +&\\
\GO{\frac{m}{2}r_1^{\omega-1}} + \GO{\frac{m}{2}(n-r_1)r_1^{\omega-2}} & \text{ if } 0<r_1< n.\\
\end{array}
\right.
$$

Consider the $i$th recursive level: the matrix is split row-wise into
$2^i$ slices of row dimension $m/2^i$.
We denote by $r_j^{(i)}$ the
rank of each of these slices, indexed by $j=0\dots 2^i-1$. 
For example $r_{1}^{(i)} = \text{rank}(\submatrix{A}{0}{\frac{m}{2^i}}{0}{n})$.

At the $i$th recursive level 
the total number of field operations is in
$$
O\left (
\frac{m}{2^i} \sum_{j=0}^{2^{i-1}}\left(r_{2j+1}^{(i)}\right)^{\omega-1} +
\frac{m}{2^i} n \sum_{j=0}^{2^{i-1}}\left(r_{2j+1}^{(i)}\right)^{\omega-2}
\right ).
$$

Since $\omega \geq 2$ we have
$$\sum_{j=0}^{2^{i-1}}\left(r_{2j+1}^{(i)}\right)^{\omega-1} \leq
r^{\omega-1}.$$
Now, using the fact that 
$
a^{\omega-2}+b^{\omega-2} \leq 2^{3-\omega} (a+b)^{\omega-2}\  \text{ for }
2<\omega\leq 3,
$
we also have
$$
\sum_{j=0}^{2^{i-1}}\left(r_{2j+1}^{(i)}\right)^{\omega-2} \leq
\left(2^{i-1}\right)^{3-\omega} r^{\omega-2}.
$$

Therefore, we obtain
\begin{eqnarray*}
T_\cupd(m,n,r) &=& \left (\sum_{i=1}^{\log_2 m} \frac{m}{2^i} r^{\omega-1}
  + \sum_{i=1}^{\log_2 m}nm\left(\frac{r}{2^i}\right)^{\omega-2}\right ) ,
\end{eqnarray*}
which for $\omega>2$ is in $\GO{mnr^{\omega-2}}$.
\qed
\end{pf}

We refer to~\ref{app:step-form} for a discussion on why the
\texttt{StepForm} algorithm does not have a rank sensitive time complexity.

\subsection{Space complexity}\label{ssec:memory}

In the presentations of Algorithms~\ref{alg:trsm}---\ref{alg:cup} we
exhibited the fact that no temporary storage was used.  Consequently
all of these algorithms, as well as  \texttt{RedEchelon}
(Algorithm~\ref{alg:redcolechelon}), work in-place as per
Definition~\ref{def:inplace}.  For a square and nonsingular input matrix,
\texttt{RedEchelon} thus gives an in-place algorithm to compute the
inverse.

For comparison, the \texttt{GaussJordan} algorithm involves  products
of the type $C\leftarrow A\times C$ (Algorithm~\ref{alg:gauss-jordan},
lines~\ref{gaussj:step:tempmm1} and~\ref{gaussj:step:tempmm2}),
which requires a copy of the input matrix $C$ into a temporary
storage in order to use a usual matrix multiplication algorithm.
These matrix multiplication in lines~\ref{gaussj:step:tempmm1}
and~\ref{gaussj:step:tempmm2} could be done in-place using
Algorithm~\ref{alg:inplacemm}, but the leading constant in the time
complexity of this version of \texttt{GaussJordan} increases to $3+1/3$
from 2 for $\omega =3$.

\begin{algorithm}
  \DontPrintSemicolon 
  \KwData{$A$ an $m \times m$ matrix over a field.}
  \KwData{$B$ an $m \times n$ matrix over a field.}
  \KwEnsures{$B\leftarrow A\times B$.} 
\Begin{
      $(r,[i_0,\dots, i_{r-1}],P) \leftarrow \cupd
      (A)$\tcc*{$A\leftarrow [C\backslash U]$} $B\leftarrow PB$\;
      $\trmm(\text{Left},\text{Up},\text{Unit},A,B)$\tcc*{$B\leftarrow
      UB$}
      $\trmm(\text{Left},\text{Low},\text{NonUnit},A,B)$\tcc*{$B\leftarrow
      CB$} $\trlum(A)$\tcc*{$A\longleftarrow CU$} $A\longleftarrow AP$\;
    }
\caption{$\texttt{InPlaceMM}(A,B)$} \label{alg:inplacemm}
 \end{algorithm}

We refer to~\ref{app:step-form} for a discussion on why the
\texttt{StepForm} algorithm is not in-place.

\subsection{Leading constants}

For a finer comparison we compute the constant of the leading term in
the complexities of all algorithms presented previously.  For simplicity
we assume that $m=n=r$ (i.e., the input matrix is square and
nonsingular).   
The complexity of each algorithm is then of the form $K_\omega n^{\omega}
+ o(n^{\omega})$ for some leading constant $K_{\omega}$ that is a function 
of the particular
exponent $\omega$ and corresponding constant $C_\omega$ 
such that two $n \times n$ matrices can be multiplied together
in time $C_\omega n^{\omega} + o(n^{\omega})$.  
To find the leading constant $K_{\omega}$ of an algorithm
we substituted $T(n)=K_\omega
n^\omega$ into the recurrence relation for the running time.
The results are summarized in Table~\ref{tab:constants},
which also gives the numerical values of $K_\omega$ for the
choices $(\omega,C_\omega) = (3,2)$ and $(\omega,C_\omega)=(\log_2
7,6)$, corresponding respectively to classical matrix multiplication
and to Strassen-Winograd's algorithm~\citep{Wino71}.
Table~\ref{tab:constants} shows how the various rank-profile revealing
elimination algorithms range in terms of time complexity: \cupd is in
$2/3n^3+ o(n^3)$, transformation to echelon form is in $n^3+o(n^3)$, and
transformation to reduced echelon form is in $2n^3+ o(n^3)$.

\begin{figure}[htbp]
  \includegraphics[width=\textwidth]{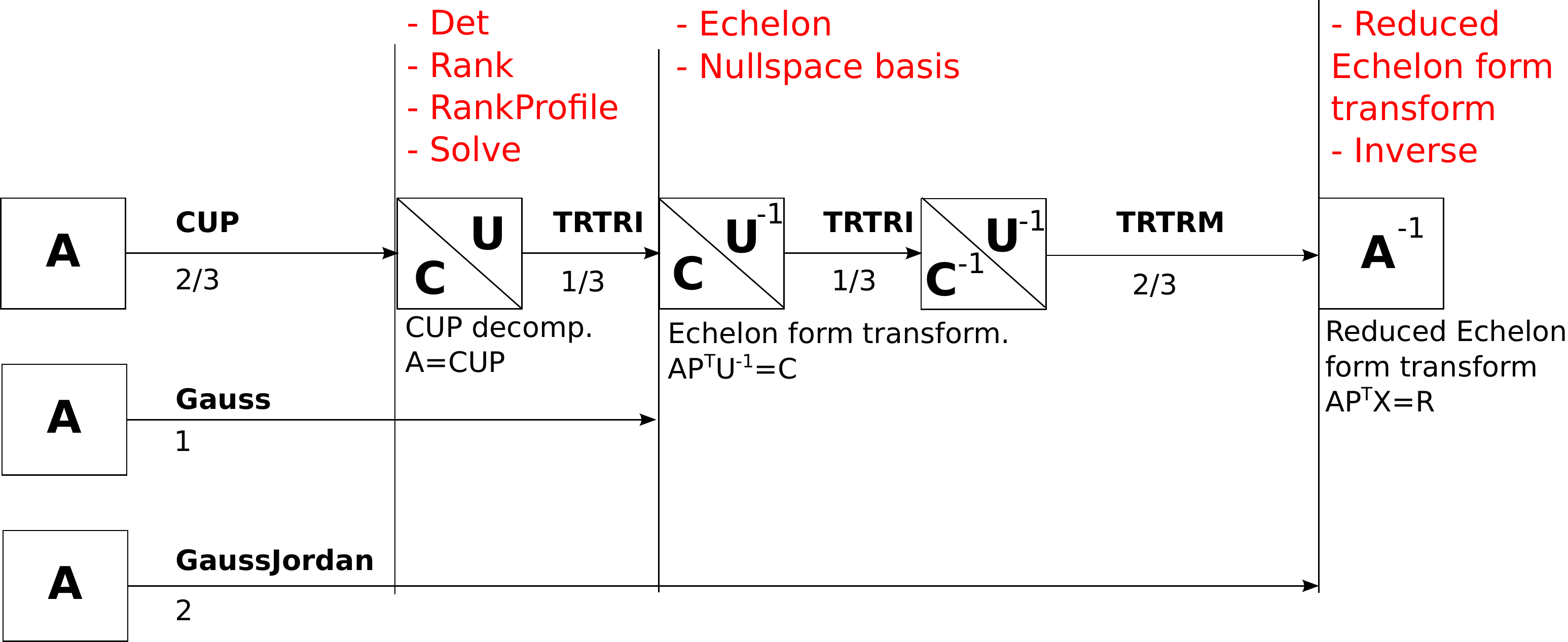}
  \caption{Application of rank revealing Gaussian elimination to linear algebra
problems.}
\label{fig:cte}
\end{figure}

Figure~\ref{fig:cte} summarizes the comparison between the three
approaches
\begin{itemize}
\item
\texttt{CUP} $\rightarrow$ \colechelon $\rightarrow$ \redcolechelon, 
\item \gauss, and 
\item \gaussjordan
\end{itemize}
with respect to their application to solving various classical
linear algebra problems.

The following observations can be made:
\begin{itemize}
\item Algorithm \cupd is sufficient (i.e., the best choice in terms
of time complexity) for
computing the determinant, the rank, the rank profile, and the 
solution of a linear
system: all of these invariants can be computed in time
$K_{\omega}n^{\omega} + o(n^{\omega})$ where $K_{\omega}$
is the leading constant for Algorithm \texttt{CUP},
for example $K_{\omega}=2/3$ in the case where $\omega=3$.
\item Computing a transformation to echelon form (and thus a basis of the
nullspace) is done by Algorithms \colechelon and \gauss
with the same time complexity. In particular, as indicated in
Table~\ref{tab:constants}, the leading coefficient $K_{\omega}$
in the complexities of these algorithms is the same.
\item Computing a transformation to reduced echelon form can be
done using Algorithms \redcolechelon or \gaussjordan.  Morever, since the
reduced echelon form of a nonsingular matrix is the identity matrix and
the corresponding transformation matrix is the inverse, these algorithms
are also algorithms for matrix inversion.  The two algorithms have the
same leading coefficient in the 
time complexity for $\omega=3$ (namely, $2n^3$), but for $\omega =\log_2 7$
Algorithm~\gaussjordan is 10\% faster.
\end{itemize}
We conclude that the \cupd decomposition is the algorithm of choice
to implement rank profile revealing Gaussian elimination: it allows to
compute all solutions in the best time complexity (except the case of
the inverse with $\omega=\log 7$, where Algorithm \gaussjordan is faster),
either directly or using additional
side computations.  However, note that \gaussjordan is not in-place; applying
the technique of Algorithm~\ref{alg:inplacemm} to make \gaussjordan in-place
increases the constant from 8 to about 21 in the case where $\omega=\log_2 7$.

\begin{landscape}
\begin{table}[htbp]
  \begin{center}
      \renewcommand{\arraystretch}{1.3}
    \begin{tabular}{cccccc}
    \toprule
    Algorithm & Operation & Constant $K_\omega$ & $K_3$ & $K_{\log_27}$&in-place\\
    \midrule
    \mm & $C\leftarrow AB$ & $C_\omega$& 2 & 6 & $\times$\\
    \trsm& $B\leftarrow BU^{-1}$ & $\frac{1}{2^{\omega-1}-2}C_\omega$ & $1$  & $4$&$\vee$\\
    \trtri& $U\leftarrow U^{-1}$ &
    $\frac{1}{(2^{\omega-1}-2)(2^{\omega-1}-1)}C_\omega$&$ 0.33$&$1.6$&$\vee$\\
    \begin{tabular}{c}
      \trulm,\trlum, \cupd
    \end{tabular}
    & 
      \footnotesize $\begin{bmatrix} L\backslash U \end{bmatrix} \leftarrow UL,LU ; 
    A\leftarrow \begin{bmatrix} C\backslash U \end{bmatrix}$ 
    &
    $\left(\frac{1}{2^{\omega-1}-2}-\frac{1}{2^{\omega}-2}\right)C_\omega$ &
    $ 0.66$ & $2.8$& $\vee$ \\
    \colechelon, \texttt{Gauss}& $A \leftarrow \begin{bmatrix}C\backslash U^{-1}\end{bmatrix}$
      &$\left(\frac{1}{2^{\omega-2}-1}-\frac{3}{2^{\omega}-2}\right)C_\omega$ & 1 &
      $4.4$&$\vee$\\
    \redcolechelon& $A \leftarrow \begin{bmatrix}R\backslash T\end{bmatrix}$ &$\frac{2^{\omega-1}+2}{(2^{\omega-1}-2)(2^{\omega-1}-1)}C_\omega$& 2 &
    $8.8$ & $\vee$\\
      \texttt{GaussJordan} 
      &$A \leftarrow \begin{bmatrix}R\backslash T\end{bmatrix}$&$\frac{1}{2^{\omega-2}-1}C_\omega$&2&8&$\times$\\
      \texttt{StepForm} 
      &$A \leftarrow \begin{bmatrix}C\backslash U^{-1}\end{bmatrix}$&$\left(\frac{5}{2^{\omega-1}-1}+\frac{1}{(2^{\omega-1}-1)(2^{\omega-2}-1)}\right)C_\omega$&4&$15.2$&$\times$\\
    \bottomrule
    \end{tabular}
    \caption{Constants of the leading term $K_\omega n^\omega$ in the algebraic complexities
    for $n\times n$ invertible matrices.}
  \label{tab:constants}
  \end{center}
\end{table}
\end{landscape}

\section{Row echelon form and the \texttt{PLE} decomposition} \label{sec:rowech}

All the algorithms and decompositions of the previous sections deal
with column echelon forms that reveal the row rank profile of the
input matrix.  By matrix transposition, similar results can be obtained
for row echelon forms and column rank profile.
The natural analogue to our central tool, the \cupd decomposition,
is the \ple decomposition: a tuple $(P,L,E)$ such that $(E^T,L^T, P^T)$
is a \cupd decomposition of the transposed input matrix $A^T$:
$$
 \overset{P^TA}{
\begin{bmatrix}
     *&*&*&*&*&*&*\\
     *&*&*&*&*&*&*\\
     *&*&*&*&*&*&*\\
     *&*&*&*&*&*&*\\
     *&*&*&*&*&*&*\\
   \end{bmatrix}}
=
 \overset{L}{\begin{bmatrix}
     1&&&\\
     *&1&\\
     *&*&1\\
     *&*&*&1 \\
     *&*&*& &1 \\
\end{bmatrix}}
 \overset{E}{\begin{bmatrix}
     \overline{*_1}&*&*&*&*&*&*\\
     &&&\overline{*_2}&*&*&*\\
     &&&&\overline{*_3}&*&*\\
     \\
     \\
   \end{bmatrix}.}
$$
To compute a \ple decomposition one could apply
Algorithm~\ref{alg:cup} with pre- and post-transpositions.  A better
option 
is to derive a ``transposed'' version of Algorithm~\ref{alg:cup},
which directly computes a \ple decomposition in-place; 
as Algorithm~\ref{alg:ple} shows, this is immediate
to achieve, and is what is implemented in the \texttt{M4RI}
library~\citep{M4RI} for dense linear algebra over $\text{GF}(2)$.
The compact rowmajor storage together with the applications driving
its development (namely Gr\"obner basis computations~\citep{F99a})
 imposed the row echelon as a standard.
The \texttt{FFLAS-FFPACK}~\citep{FFLASFFPACK:lib}
library for dense linear algebra over word-size finite fields
implements both the \cupd and \ple Algorithms.

\newpage 

\begin{algorithm}
  \DontPrintSemicolon
  \KwData{$A$ an $m \times n$ matrix.}
  \KwResult{$(r,[i_0,\dots,i_{r-1}],P)$.}
  \KwEnsures{$A \leftarrow [L\backslash E]$ such that 
    embedding $L$ in an $m\times m$ unit lower triangular matrix makes $A=PLE$ a \ple decomposition of $A$, $r=\textnormal{rank}(A)$ and
    $[i_0,\dots,i_{r-1}]$ its column rank profile.}
  \Begin{
      \If{$n=1$}{
        \If{ $A\neq 0$}{
          $i \leftarrow $ row index of the first nonzero entry of $A$\;
          $P\leftarrow T_{0,i}$, the transposition of indices $0$ and $i$\;
          $A\leftarrow P A$\;
          \lFor{$i\leftarrow 2\dots m$}{$A_{i,0}\leftarrow A_{i,0}A_{0,0}^{-1}$\;}
          \Return{$(1, [0], P)$}
        }{
          \Return{$(0, [], I_n)$}
        }
      }
      $k \leftarrow \lfloor \frac{n}{2} \rfloor$
      \tcc*{$A=
        \left[
        \begin{array}{c|c}
          A_1&A_2
        \end{array}\right]$
      }
      \lnl{ple:step:rec1}$(r_1,[i_0,\dots,i_{r_1-1}],P_1) \leftarrow \ple(\submatrix{A}{0}{m}{0}{k})$ 
         \tcc*{$A_1 \leftarrow P_1 \begin{bmatrix} L_1\\M_1 \end{bmatrix} E_1$}
         \If{$r_1 = 0$}{
           $(r_2,[i_0,\dots,i_{r_2-1}],P_2)\leftarrow \ple(\submatrix{A}{0}{m}{k}{n})$\;
           \Return{$(r_2,[i_0+k,\dots,i_{r_2-1}+k],P_2)$}  
         } 
         $\submatrix{A}{0}{m}{k}{n} \leftarrow P_1^T\submatrix{A}{0}{m}{k}{n} $\;
         \tcc{\small $A=
           \begin{bmatrix}
             L_1\backslash E_1 & A_{12} \\
             \begin{array}{cc}
               M_1&0
             \end{array}
              & A_{22}
           \end{bmatrix}
           $ with $L_1\backslash E_1$ $r_1\times k$, and $M_1$  $(n-k)\times r_1$}
         \lnl{ple:step:trsm} $\trsm(\text{Left},\text{Low},\text{Unit},\submatrix{A}{0}{r_1}{0}{r_1},\submatrix{A}{0}{r_1}{k}{n}$)
         \tcc*{$G \leftarrow L_1^{-1}A_{12}$}
         \If{$r_1=m$}{
            \Return {$(r_1,[i_0,\dots,i_{r_1-1}],P_1)$}
         }
         \lnl{ple:step:mm}$\mm(\submatrix{A}{r_1}{m}{0}{r_1} , \submatrix{A}{0}{r_1}{k}{n},\submatrix{A}{r_1}{m}{ k}{n},-1,1)$
	 \tcc*{$H \leftarrow A_{22} - M_1G$}
	 \lnl{ple:step:rec2}$(r_2,[j_0,\dots,j_{r_2-1}],P_2) \leftarrow \ple(\submatrix{A}{r_1}{m}{k}{n})$\;
         $\submatrix{A}{r_1}{m}{0}{r_1} \leftarrow P_2^T \submatrix{A}{r_1}{m}{0}{r_1}  $
         \tcc*{$N_1\leftarrow P_2^TM_1$}

         \lnl{ple:step:permute}
         \For{$j\leftarrow r_1\dots r_1+r_2$}{
           
           $\colofmatrix{A}{j}{m}{j} \leftarrow \colofmatrix{A}{j}{m}{j+k-r_1}$
           \tcc*{Moving $L_2$ left next to $M_1$}
           $ \colofmatrix{A}{j}{m}{j+k-r_1} \leftarrow [0]$\;
         }
         $P \leftarrow P_1\text{Diag}(I_{r_1}, P_2) $\;
         \Return {$ (r_1+r_2,[i_0,\dots,i_{r_1-1},j_0+k,\dots,j_{r_2-1}+k], P)$}
    }
    \caption{$\ple(A)$}
    \label{alg:ple}
  \end{algorithm}

\newpage

\section{Conclusion} \label{sec:conclusion}

We have presented an algorithm computing the rank profile revealing 
\cupd decomposition of
a matrix over a field.  We have shown that the algorithm enjoys the following
features:
\begin{enumerate}
\item The algorithm reduces
     to matrix-matrix multiplications and therefore has a subcubic 
    time complexity.
\item The complexity is rank sensitive
    of the form  $O(mnr^{\omega-2})$, where $r$ is the rank of the
   $m \times n$ input matrix.
\item The algorithm is in-place, that is,  only $O(1)$
    extra memory allocation for field elements is required beyond 
    what is needed for the matrix products.
\item Used as a building block for most common computations in linear algebra,
    the algorithm 
    achieves the best constant in the leading term of the time
    complexity. The only exception is for the reduced echelon form with
    $\omega=\log_2 7$, where the constant is $8.8$ instead of $8$ 
    for Algorithm \texttt{GaussJordan}.
\end{enumerate}
Among the set of Gaussian elimination algorithms studied here, 
the algorithm for \cupd decomposition is the only
one satisfying all the above conditions. 
For these reasons it has been chosen for the implementation of Gaussian
elimination in the \texttt{FFLAS-FFPACK} and
\texttt{LinBox} libraries for dense matrices over word-size
finite fields~\citep{FFLASFFPACK:lib, LinBox:lib}, as well as for the \texttt{M4RI}
and~\texttt{M4RIE} libraries for dense matrices over $\text{GF}(2)$
and $\text{GF}(2^e)$, respectively~\citep{M4RI,Alb11,AlbBarPer11}.

\appendix

\section{The \texttt{GaussJordan} algorithm}
\label{app:storj}

The \texttt{GaussJordan} algorithm~\citep[Algorithm 2.8]{Sto00} was
originally presented as a fraction free algorithm for transforming a
matrix over an integral domain to reduced row echelon form.  We
give here a simpler version of the algorithm over a field.  
The principle is to transform
the matrix to reduced row echelon form, slice by slice. The general
recursive step reduces a contiguous slice of columns of width $w$
starting at column $k$ by
\begin{enumerate}
\item recursively reducing the first half of the slice of columns (of width $w/2$),
\item updating the second half of the slice of columns (also of width $w/2$),
\item recursively reducing the second half,
\item composing the two transformation matrices.
\end{enumerate}
The algorithm is described in full details in Algorithm~\ref{alg:gauss-jordan}. 
Calling \texttt{Gauss\-Jordan}\-$(A,0,0,n)$ computes the reduced row echelon form
and the associated  transformation matrix.
Once
again, the presentation based on the indexing of the submatrices shows where
all matrices are located in order to illustrate the need for extra memory
allocation.

\begin{algorithm}
\footnotesize
\DontPrintSemicolon
  \KwData{$A$ an $m \times n$ matrix over a field.}
  \KwData{$k,s$ the column, row of the left-top coefficient of the block to reduce.}
  \KwData{$w$ the width of the block to reduce.}
  \KwResult{$(r,P,Q)$, where $P,Q$ are permutations matrices of order $m$
  and $w$.}
  \KwEnsures{$r=\text{rank}(\submatrix{A}{s}{m}{k}{k+w})$ and 
    $A\leftarrow
      \left[\begin{array}{c|c|c}
      \submatrix{A}{0}{m}{0}{k}\ &
      \begin{array}{cc}
        X_1&F\\
        X_2&G\\
        X_3&0
      \end{array}
      &      \submatrix{A}{0}{m}{k+w}{n}
      \end{array}\right] $ such that
       $       \begin{bmatrix}
         I_s & X_1 & \\
         &X_2&\\
         &X_3&I_{m-r-s}
       \end{bmatrix}
       \!\!P \submatrix{A}{0}{m}{k}{k+w} =
       \begin{bmatrix} 0& F\\  I_r&G\\ 0& 0\end{bmatrix}\!\!
       \begin{bmatrix} I_k\\&Q   \end{bmatrix} $.
  }
  \Begin{
      \eIf{$w=1$}{
        \eIf{$\colofmatrix{A}{s}{m}{k}\neq [0]$}{
          $j\leftarrow $ the row index of the first nonzero entry of $\colofmatrix{A}{s}{m}{k}$\;
          $P \leftarrow T_{s,j}$ the transposition of indices $s$ and $j$\;
           $A\leftarrow PA$\;
          \Return $(1,P,[1])$\;
        }{
         \Return $(0,I_m,[1])$\;
        }
      }{
        $h\leftarrow \lfloor w/2 \rfloor$\;
        $(r_1, P_1, Q_1) = \texttt{GaussJordan}(A, k, s, h) $\;
        \tcc{Notations: $A=
          \left[\begin{array}{c|c|c}
              *&
              \begin{array}{c|c}
                \begin{array}{cc}
                  X_1&F\\
                  X_2&E\\
                  X_3&
                \end{array}
                &
                \begin{array}{c}
                  Y_1\\Y_2\\Y_3
                \end{array}
              \end{array}
              &*
            \end{array}\right]
            $ where $X_2$ is $r_1\times r_1$.}
          $ t\leftarrow s+r_1;\ g\leftarrow k+h $\tcc*{top left indices of $Y_3$}
        \mm($\submatrix{A}{0}{s}{k}{k+r_1}, \submatrix{A}{s}{t}{g}{k+w}, \submatrix{A}{0}{s}{g}{k+w}, 1, 1$)
        \tcc*{$Y_1\leftarrow Y_1 + X_1Y_2$}
        
        $\text{temp}\leftarrow \submatrix{A}{s}{t}{g}{k+w}$
        \tcc*{a temporary is needed for $Y_2$}
        \lnl{gaussj:step:tempmm1}\mm($\submatrix{A}{s}{t}{k}{k+r_1}, \text{temp}, \submatrix{A}{s}{t}{g}{k+w}, 1, 0$)
        \tcc*{$Y_2\leftarrow  X_2Y_2$} 
        \mm($\submatrix{A}{t}{m}{k}{k+r_1}, \submatrix{A}{s}{t}{g}{k+w}, \submatrix{A}{t}{m}{g}{k+w}, 1, 1$)
        \tcc*{$Y_3\leftarrow Y_3 + X_3Y_2$}
        $(r_2, P_2, Q_2) = \texttt{GaussJordan}(A, g, t, w-h)$\; 
        \tcc{Notations: $A=
          \left[\begin{array}{c|c|c}
              *&
              \begin{array}{c|c}
                \begin{array}{cc}
                  X_1&E_1\\
                  X_2&E_2\\
                  X'_3&\\
                  X'_4
                \end{array}
                &
                \begin{array}{cc}
                  \begin{array}{cc}
                    Z_1&F_1\\
                    Z_2&F_2\\
                    Z_3&F_3\\
                    Z_4&
                  \end{array}
                \end{array}
              \end{array}
              &*
            \end{array}\right]
            $ with $Z_3$ $r_2\times r_2$}
        \mm($\submatrix{A}{0}{t}{g}{g+r_2}, \submatrix{A}{t}{t+r_2}{k}{k+r_1}, \submatrix{A}{0}{t}{k}{k+r_1}, 1, 1$)
        \tcc*{$ 
          \begin{bmatrix} X_1\\X_2 \end{bmatrix}
          \leftarrow 
          \begin{bmatrix} X_1\\X_2 \end{bmatrix} + \begin{bmatrix} Z_1\\Z_2 \end{bmatrix} X'_3$}
        $\text{temp}\leftarrow \submatrix{A}{t}{t+r_2}{k}{k+r_1}$
        \tcc*{a temporary is needed for $X'_3$}
        \lnl{gaussj:step:tempmm2}\mm($\submatrix{A}{t}{t+r_2}{g}{g+r_2}, temp,\submatrix{A}{t}{t+r_2}{k}{k+r_1}, 1, 0$)
        \tcc*{$X'_3\leftarrow Z_3X'_3$} 
        \mm($\submatrix{A}{t+r_2}{m}{g}{g+r_2}, \submatrix{A}{t}{t+r_2}{k}{k+r_1}, \submatrix{A}{t+r_2}{m}{ k}{k+r_1}, 1, 1$)
           \tcc*{$X'_4\leftarrow X'_4 + Z_4X'_3$}    
           $Q\leftarrow 
           \begin{bmatrix}
                I_{r_1}\\ 
                &&I_{h-r_1}\\
                &I_{r_2}\\
                &&&I_{w-h-r_2}
           \end{bmatrix}
           \begin{bmatrix}
           Q_1\\&Q_2
           \end{bmatrix}$\;
          \Return{$(r_1+r_2,P_2P_1,Q)$}
      }
    }
\caption{\texttt{GaussJordan}$(A,k,s,w)$}\label{alg:gauss-jordan}
\end{algorithm}

Assuming $m=n=r$, the time complexity satisfies:
\begin{eqnarray*}
T_{\texttt{GaussJordan}}(n,w) &=&
2T_{\texttt{GaussJordan}}(n,w/2)+2T_\texttt{MM}(w/2,w/2,n)
\end{eqnarray*}

Substituting the general form $T(n,w)=K_\omega nw^{\omega-1}$ into this recurrence
relation, we obtain $K_\omega = \frac{C_\omega}{2^{\omega-1}-1}$.

\section{The \texttt{StepForm} algorithm}
\label{app:step-form}

The \texttt{StepForm} algorithm, due to Sch\"onhage and Keller-Gehrig,
is described in~\citep[\S 16.5]{BuClSh97}. The algorithm proceeds by first
transforming the input matrix into an upper triangular matrix, and then
reducing it to echelon form.  Thus, the rank profile of the matrix only
appears at the last step of the algorithm and the block decomposition used
for the first step is not rank sensitive. As a consequence, the complexity
of the algorithm does not depend on the rank of the matrix. Note that
this limitation is not because of simplifying assumptions in the analysis
of the complexity, but because the algorithm is itself not rank-sensitive.

We now evaluate the constant of the leading term in the complexity of
the  \texttt{StepForm} algorithm under the simplifying asumption $m=n=r$.  The
description of Algorithms  $\Pi_1, \Pi_2$, and $\Pi_3$ used as sub-phases
can be found in~\citep[\S 16.5]{BuClSh97} and the references therein.

Let $T_1(n)$ be the complexity of Algorithm $\Pi_1$ applied to a $2n\times
n$ matrix, and let $T_2(n)$ be the complexity of Algorithm $\Pi_2$ applied to an
$n\times n$ matrix.

We have
$$
T_1 (n) = 4T_1(\frac{n}{2})+ 4\mm(\frac{n}{2},\frac{n}{2},\frac{n}{2})+\mm(n,n,n),$$
from which we deduce
$$
T_1(n)=C_\omega \frac{2^{\omega-2}+1}{2^{\omega-2}-1}n^\omega.
$$
We have
\begin{eqnarray*}
  T_2 (n) &=& 2T_2(\frac{n}{2})+T_1(\frac{n}{2})+
5\mm(\frac{n}{2},\frac{n}{2},\frac{n}{2}) +\mm(n,n,\frac{n}{2})\\
&=&2T_2(\frac{n}{2})+C_\omega \left(\frac{2^{\omega-2}+1}{2^{\omega-2}-1} +\frac{9}{2^\omega}\right)n^\omega
\end{eqnarray*}
from which we deduce
$$
T_2(n)=C_\omega \left(\frac{5}{2^{\omega-1}-1}+\frac{1}{(2^{\omega-1}-1)(2^{\omega-2}-1)}\right)n^\omega.
$$
Under our assumption $n=m=r$,  Algorithm $\Pi_3$ does
not perform any operation and the total complexity is
therefore $T_2(n)$.  For $(\omega, C_\omega)=(3,2)$ we obtain
$T_2(n)=4n^3$, and for $(\omega,C_\omega)=(\log_27,6)$ we obtain
$T_2(n)=\frac{76}{5}n^{\log_27}=15.2n^{2.81}$.

Finally, the algorithm does not specify that the transformation matrix
generated by algorithm $\Pi_2$ is lower triangular: it needs to be stored
in some additional memory space and thus the algorithm is not in-place.

\bibliographystyle{model2-names}
\bibliography{ple}







\end{document}